\documentclass[12pt,dvips]{article}
\usepackage[dvips]{graphicx}
\usepackage{graphicx}
\usepackage{amssymb}

\parskip=\itemsep               
\newcommand{\lwig}{\mbox{\,\raisebox{.3ex}
    {$<$}$\!\!\!\!\!$\raisebox{-.9ex}{$\sim$}\,}}

\newcommand{\lambdabar}{{\hbox{$\lambda_e$\kern-1.9ex\raise+0.45ex\hbox{--}
\kern+0.2ex}}}

\date{\empty}

\title{{\normalsize\rightline{DESY 02-053}\rightline{hep-ph/0205024}}
\vskip 1cm  
\bf Microscopic Black Hole Production \\
in TeV-Scale Gravity\thanks{Talk presented at XXX ITEP Winter School
of Physics, Moscow, Russia, February 2002.}
       \vspace{21mm}} 
\author{
Huitzu Tu\\[4mm] 
\textsl{Deutsches Elektronen-Synchrotron DESY, Hamburg, Germany}}

\begin{document}
\begin{titlepage} 
  \maketitle

\begin{abstract}
\noindent
Models with extra spatial dimensions and TeV-scale gravity offer the first 
opportunity to test the conjecture of black hole formation in 
trans-Planckian energy scattering with small impact parameters.
After a brief review of gravitational scattering at ultrahigh energies and 
scenarios of TeV-scale gravity, search strategies at the LHC, at the 
Pierre Auger (cosmic ray) Observatory and at the neutrino telescopes 
AMANDA/IceCube are illustrated with the simplest but nevertheless 
representative example: production of Schwarzschild black holes and their 
observation via Hawking radiation in the large extra dimension scenario.
Some more general features of the production of higher-dimensional black 
holes and/or uncertainties in the estimates are also outlined.

\end{abstract}

\thispagestyle{empty}
\end{titlepage}
\newpage \setcounter{page}{2}

\section{Introduction}

Black holes play an essential role in our approach to the quantum theory 
of gravity.
The history of quantum gravity is deeply intertwined with the problem of 
very high energy scattering.
It has been conjectured that microscopic black holes may be formed in 
particle collisions at energies higher than the Planck mass and with impact 
parameters smaller than a critical value
\cite{'tHooft:1987rb,Amati:1987wq,Verlinde:1992iu,Fabbrichesi:1994kd,Aref'eva:1995qs}.

In models with $\delta = D - 4$ extra spatial dimensions, where the 
Standard Model particles are assumed to reside on a 3-dimensional brane 
while only gravitons are allowed to propagate into the bulk, the Planck scale, 
which is the scale characterising quantum gravity, can be just beyond the 
electroweak scale
\cite{Arkani-Hamed:1998rs,Randall:1999ee}.
Within such TeV-scale gravity models, the conjecture suggests that particle 
collisions at TeV energies may result in the production of black holes of 
masses at this energy scale provided the colliding particles come close enough
\cite{Argyres:1998qn,Banks:1999gd,Aref'eva:2000bm,Emparan:2000rs,Giddings:2001ay,Emparan:2001ce}.

Due to their small masses, these microscopic black holes undergo decay 
processes rapidly. 
It is believed that these multi-dimensional black holes should 
Hawking-radiate \cite{Hawking:1974rv}
mainly into Standard Model particles on the brane rather than into the bulk
\cite{Emparan:2000rs}.
Thus direct observations of such black hole events are possible.
Estimates show that, depending on the value of the higher-dimensional
fundamental Planck scale, the Large Hadron Collider (LHC) may either turn 
into a black hole factory
\cite{Giddings:2002bu,Dimopoulos:2001hw}, 
where the black hole formation conjecture, the Hawking radiation law and the 
existence of extra spatial dimensions can be verified experimentally, 
or be able to put constraints on the model parameters from non-observation.
On the other hand, it is well known that particle astrophysics experiments 
are complementary to collider searches for new physics beyond the 
Standard Model. 
In the case of black hole production in TeV-scale gravity models, one finds 
\cite{Feng:2002ib,Anchordoqui:2002ei,Uehara:2001yk,Ringwald:2002vk,Kowalski:2002gb,Alvarez-Muniz:2002ga} 
that depending on the fluxes of the ultrahigh energy cosmic neutrinos, 
cosmic ray facilities such as Auger and neutrino telescopes like AMANDA 
and RICE may have an opportunity to see the first sign or put constraints 
on black hole production parameters before LHC starts operating. 
IceCube has even discovery potential beyond the LHC reach.

In section 2 the general aspects of gravitational scattering at ultrahigh 
energies will be briefly reviewed.
Two scenarios of extra dimension models in which TeV-scale gravity can be 
realised are outlined in section 3.
Section 4 will be devoted to black hole production and evaporation in 
TeV-scale gravity, with some remarks concerning the working hypothesis 
adopted.  
In section 5 we illustrate search strategies at colliders, cosmic ray 
facilities and neutrino telescopes with the simplest, nevertheless 
representative example: production and observation via Hawking radiation 
of the microscopic Schwarzschild black holes in the large extra dimension 
scenario.

\section{Trans-Planckian energy scattering}

In this section, we briefly review the current understanding of particle 
scattering at ultrahigh energies in conventional 4-dimensional theories 
\cite{'tHooft:1987rb,Amati:1987wq,Kabat:1992tb,Verlinde:1992iu,Fabbrichesi:1994kd,Aref'eva:1995qs}.
We shall present arguments why particle scattering at trans-Planckian
energies can be treated semiclassically and why one expects black hole
formation for small enough impact parameters.

We start by recalling the two crucial dimensionless parameters which
characterise gravitational 
scattering\footnote{Throughout this paper we set $c = 1$ but keep $\hbar$
in this section.}
\begin{equation}
   \alpha_G \equiv \frac{G_{\rm N}\, s}{\hbar} \hspace{0.5cm} 
   {\rm and} \hspace{0.5cm} \frac{G_{\rm N}\, s}{J} = 
   \frac{R_{\rm S}}{b}\, .
\label{dimless-para}
\end{equation}
Here, $G_{\rm N}$ is Newton's constant, $\sqrt{s}$ is the center-of-mass
energy, $b$ the impact parameter and $J = (b / 2)~\sqrt{s}$ the total 
angular momentum.
The Schwarzschild radius associated with a certain centre-of-mass energy,
\begin{equation}
   R_{\rm S} \equiv 2\, G_N \sqrt{s}\, ,
\end{equation}
is the length scale at which curvature effects become significant.
Other relevant length scales in ultrahigh energy scattering regime include 
the Planck length $\lambda_{\rm Pl} = \sqrt{G_{\rm N} \hbar}$, which 
signifies when quantum gravity effects become important, and the 
string length scale,
\begin{equation}
   \lambda_{\rm str} = \sqrt{\alpha' \hbar}
   = g^{-1}_{\rm str}\, \lambda_{\rm Pl}\, .
\end{equation} 
It is the fundamental length scale when the problem is studied within 
the context of string theory.
Here $\alpha'$ is the Regge slope and $g_{\rm str}$ is the gauge coupling 
at the string scale.
Conventionally, it is of order $g^2_{\rm str} / (4 \pi) \sim 0.04$.
In the remainder of this section, we shall assume 
\begin{equation}
   R_{\rm S} \gg \lambda_{\rm str}\, .
\end{equation}
In this regime, the internal structure of the strings can be ignored and 
it is sufficient to employ low energy effective field theory.
So one is indeed left with the two dimensionless parameters 
eq.~(\ref{dimless-para}) characterising the collision.

The parameter $\alpha_G$ in eq.~(\ref{dimless-para}) is the gravitational 
equivalent of the fine-structure constant.
The essential difference lies in its explicit energy dependence, which is 
characteristic for gravitational interactions since energy itself plays 
the role of gravitational charge. 
Trans-Planckian energy scattering is defined by 
\begin{equation}
   \alpha_G = \frac{G_{\rm N}\, s}{\hbar} \gg 1 
   \Leftrightarrow \sqrt{s} \gg M_{\rm Pl}\, ,
\end{equation}
with $M_{\rm Pl} = (G_{\rm N} / \hbar)^{-1 / 2} \simeq 
1.2 \cdot 10^{19}~{\rm GeV}$ the Planck mass. 
In this regime, gravitational interactions dominate over other gauge 
interactions. 
The gravitational scattering process in this regime is semiclassical and 
calculable by non-perturbative approaches only.

Note however that the second crucial parameter $R_{\rm S} / b$ in 
eq.~(\ref{dimless-para}) can still be varied at will.
While trans-Planckian energy scattering is always a semiclassical process, 
very different physics is expected to emerge depending on the value of 
$R_{\rm S} / b$.

In the small scattering angle limit, $- t / s \ll 1$ 
($t$ is the Lorentz-invariant momentum transfer), 
or, correspondingly, for large impact parameters $R_{\rm S} / b \ll 1$,
the gravitational scattering cross section has been calculated by various 
semiclassical approaches
\cite{'tHooft:1987rb,Amati:1987wq,Kabat:1992tb,Verlinde:1992iu,Fabbrichesi:1994kd,Aref'eva:1995qs},
e.g. in Ref.~\cite{'tHooft:1987rb} by solving the Klein-Gordon equation 
for a particle encountering the gravitational shock wave
generated by another a fast-moving particle
\cite{Aichelburg:1971dh},
by employing the eikonal approximation 
\cite{eikonal:1969}
for four-dimensional gravity 
\cite{Kabat:1992tb}
and for string theory
\cite{Amati:1987wq},
or by describing quantum gravitational effects by means of a
topological field theory
\cite{Verlinde:1992iu}.
All calculations agree with the claim that the process is dominated 
by eikonalised single graviton exchange.

As for smaller impact parameters $b$, it is conjectured that black hole 
formation should set in below a critical value of $b / R_{\rm S}$ 
of order one
\cite{'tHooft:1987rb,Amati:1987wq,Kabat:1992tb,Verlinde:1992iu,Fabbrichesi:1994kd,Aref'eva:1995qs}. 
Are there sufficient evidences for one to believe in this conjecture?

In $(2 + 1)$--dimensional Anti-de Sitter space it has already been shown 
that a 3--dimensional black hole \cite{Banados:1992wn}
is inevitably created in head-on particle collisions above some energy 
threshold \cite{Matschull:1999rv}.
In four-dimensions, Penrose \cite{Penrose:1974} and D'Eath and Payne
\cite{D'Eath:1992hb} (see also Ref.~\cite{D'Eath:1996nf})
studied the collision of two black holes moving at nearly the speed of light, 
and showed that closed trapped surfaces can be formed for a range of impact 
parameters.
This study is redone and extended to particle collisions in higher dimensions 
recently \cite{Eardley:2002re}
by following the same approach, namely by modelling the gravitational field 
of each incoming particle by a plane-fronted gravitational shock wave. 
Also in this case, closed trapped surfaces were found.
Furthermore, results from numerical studies on gravitational collapse of 
rotating stars also show black hole formation below a critical angular 
momentum $J$ \cite{Stark:it}.
One may consider this as another supporting argument.
For further attempts to prove the black hole formation conjecture, see 
Refs.~\cite{Solodukhin:2002ui,Kohlprath:2002yh}.

\section{TeV-scale gravity}

Within models with extra spatial dimensions, the scale characterising
quantum gravity can actually be in the TeV range. 
This puts the remote possibility of microscopic black hole production 
within phenomenological reach.
In this section we shall shortly outline two possible constructions 
of TeV-scale gravity models: 
the large compact extra dimension scenario \cite{Arkani-Hamed:1998rs}
and the warped scenario \cite{Randall:1999ee}.
(For a review of extra dimensions see e.g. \cite{Rubakov:2001kp}).

In the large extra dimension scenario, it is assumed that gravitons are 
allowed to propagate freely into the bulk, while other Standard Model fields 
must be localised on a 3--brane.
The 4--dimensional effective action describing long-distance gravity is then 
obtained by integrating the $D = (\delta + 4)$--dimensional gravitational 
action over the extra dimension coordinates, of which the metric is assumed 
to be independent
\begin{equation}
   S = \frac{- 1}{16 \pi G_D} \int d^D X \sqrt{- g^{(D)}}{\cal R}^{(D)}  
   \hspace{0.1cm} \Rightarrow \hspace{0.1cm} 
    S_{\rm eff} = \frac{- V_{\delta}}{16 \pi G_D} 
   \int d^4 x \sqrt{- g^{(4)}}{\cal R}^{(4)}\, . 
\end{equation}
Therefore,
\begin{equation}
   \frac{V_{\delta}}{G_D} = \frac{1}{G_{\rm N}}\, , \hspace{0.5cm}
   {\rm with} \hspace{0.3cm}
   G_D = \frac{(2 \pi)^{\delta - 1} \hbar^{\delta + 1}}{
   4 M^{\delta +2}_D}
\end{equation}
relating the $D$-dimensional Newton's constant to the fundamental Planck 
scale $M_D$, in the convention of \cite{Giudice:2001ce}.

The hierarchy between the four-dimensional Planck mass $M_{\rm Pl}$ and 
the fundamental Planck scale $M_D$ arises then from the large volume of 
the extra dimensions
\begin{equation}
   M^2_{\rm Pl} = \frac{4}{(2 \pi)^{\delta - 1}}~V_{\delta}
   ~M^{2 + \delta}_D\, .
\end{equation}
For a fundamental Planck scale $M_D \sim$ TeV, sizes of the compact extra 
dimensions are required to be (in the case that all extra dimensions are 
of the same order) 
\begin{equation}
   r_c \sim \frac{1}{M_D} \left(\frac{M_{\rm Pl}}{M_D} \right)^
   {2 / \delta} \sim 10^{\frac{32}{\delta} - 17} \cdot 
   \left(\frac{1~\rm {TeV}}{M_D}\right)^{1 + 2 / \delta}~{\rm cm}\, ,
\label{compactification-radii}
\end{equation}
and can even be as large as $\sim 0.1~{\rm mm}$, since this is the smallest 
distance ever probed directly so far.
One sees also that $\delta = 1$ is immediately ruled out by astronomical 
data, while for $\delta = 2$ to 6 the compactification radii $r_c$ range 
from a sub-millimeter to a few fermi.

Warped scenarios provide an alternative approach to explain the hierarchy:
the weak scale is generated from the Planck mass through the
``warp factor'', which is an exponential function of the compactification 
radius, arising not from gauge interactions but from the background metric.

A simple explicit example of this mechanism is demonstrated in
Ref.~\cite{Randall:1999ee} 
with two 3-branes and one compact extra dimension, and with the Standard 
Model fields residing on the ``visible'' brane.
The 3-branes, extending in the $x^{\mu}$ directions, are located at 
$0$ and $r_c\, \pi$ in the fifth dimension.
The effects of the branes on the bulk gravitational metric are taken into 
account, and are expressed by the ``vacuum energy'' terms $V_{\rm vis}$ 
and $V_{\rm hid}$ in the action   
\begin{eqnarray}
  S &=& S_{\rm gravity} + S_{\rm vis} + S_{\rm hid}\, ,
  \hspace{0.4cm} S_{\rm gravity} = \int d^4 x ~\int^{\pi}_{- \pi}
  d \phi\, \sqrt{- g^{(5)}}\, \{ - \Lambda + 2\, 
  M^3\, {\cal R}^{(5)} \}\, , \nonumber \\
  S_{\rm vis} &=& \int d^4 x\, \sqrt{- g_{\rm vis}}\, 
  \{ {\cal L}_{\rm vis} - V_{\rm vis} \}\, ,  
  \hspace{0.4cm}
  S_{\rm hid} = \int d^4 x\, \sqrt{- g_{\rm hid}}\, 
  \{ {\cal L}_{\rm hid} - V_{\rm hid} \}\, ,
\end{eqnarray}
where $\Lambda$ is the cosmological term and $M$ the Planck scale in the 
five-dimensional spacetime.
The solution to the five-dimensional Einstein's equations is required to 
respect four-dimensional Poincar${\rm \acute e}$ invariance in the 
$x^{\mu}$ directions, and is found to be
\begin{equation}
   d s^2 = e^{- 2\, k\, r_c\, \vert \phi \vert}\,
   \eta\,_{\mu\, \nu}\, d x^{\mu}\, d x^{\nu} + r^2_c\, 
   d\, \phi^2,
\end{equation}
where $k$ is the single scale which fine-tunes between 
$\Lambda$, $V_{\rm vis}$ and $V_{\rm hid}$.
The warp factor contained in the metric generates an exponential 
hierarchy between $M_{\rm Pl}$ and the electroweak scale $M_{\rm EW}$: 
\begin{equation}
   M_{\rm EW} \sim e^{- k\, r_c \pi}\, M_{\rm Pl}.
\end{equation}
To account for the observed 15 orders of magnitude between $M_{\rm Pl}$ 
and TeV physical mass scales, only $k\, r_c \approx 10$ need to be 
required.
Therefore, other than in the large extra dimension scenarios, the size of 
the extra dimension need not be much larger than the natural length scale 
$M^{-1}_{\rm Pl}$.
Furthermore, the largeness of the Planck mass $M_{\rm Pl}$ or the weakness 
of the gravity for an observer on the visible brane can be interpreted 
as due to the small overlap of the graviton wave functions in 5 dimensions 
with the visible brane.

Phenomenology of extra dimension scenarios includes deviation from 
Newtonian gravity in distances smaller than the compactification radii, 
Kaluza-Klein graviton emission and exchange\footnote{For an exhaustive
list of references in this context, see also Ref.~\cite{Ringwald:2002vk}.}, 
and the possible production of microscopic black holes in trans-Planckian 
energy scattering.
Lower limits on the fundamental Planck scale $M_D$ in the large extra 
dimension scenario are derived from sub-millimeter tests of Newton's law
by Cavendish-type experiments 
\cite{Adelberger:2002ic}, 
from collider searches for graviton 
emission and exchange in scattering processes
\cite{Abe:2001nq}, 
as well as from astrophysical and cosmological considerations, such as 
the rate of supernova cooling and the diffuse $\gamma$--ray spectrum
\cite{Hannestad:2002xi}.
Some of the lower limits are listed in Table~\ref{LED}.

\begin{table}
\begin{center}
\begin{tabular}{|l|ccc|}
\hline \hline
      & $M_6$ & $M_8$ & $M_{10}$ \\ 
\hline 
  LEP II 
 & $1.45$ &  $0.87$ & $0.61$\\
astroph.    & 1881 
                 & 8.8 & 0.61 \\ 
\hline \hline
      & $M_6$ & $M_7$ & $M_8$ \\ \hline
LHC & $4.0\div 8.9$ &  $4.5\div 6.8$ & $5.0\div 5.8$\\ 
\hline \hline
\end{tabular}
\vspace{0.3cm}
\caption{\small
$95\%$ C.L. lower limits on the fundamental Planck scale $M_D$ (in TeV)
in the large extra dimension scenario \cite{Arkani-Hamed:1998rs}, 
derived from the searches for Kaluza-Klein graviton emission in the 
$e^+ e^- \rightarrow \gamma G$ channel by the L3 Collaboration
\cite{Abe:2001nq}, and limits from astrophysical considerations
\cite{Hannestad:2002xi}.
Expected lower limits on $M_D$ (in TeV) for $D = 6, 7, 8$ from the searches 
for graviton emission in the $p p \rightarrow g G$ channel at the LHC 
\cite{Abe:2001nq} are also listed.
}
\label{LED}
\end{center}
\end{table}

To summarise, a fundamental Planck scale as low as $M_D \sim {\cal O} (1)$ 
TeV is still allowed for $\delta \geq 4$ flat or $\delta \geq 1$ warped 
extra dimensions. 
More about the experimental bounds on extra dimension models can be found 
in e.g. Ref.~\cite{LED:limits}.

\section{Black hole production and evaporation in TeV-scale gravity}

The old conjecture of black hole formation in trans-Planckian energy
particle collisions has attracted again a great deal of attention,
thanks to the interesting proposal that the fundamental Planck scale lies 
just beyond the electroweak scale.
In this section we introduce our working hypothesis for black hole 
production and evaporation in the context of TeV-scale gravity.
Some caveats/uncertainties in the estimates are also pointed out.

\subsection{Black hole production}

We start again by specifying the regime of our interests and recalling 
the relevant length scales therein.
In TeV-scale gravity, the trans-Planckian energy regime corresponds to 
\begin{equation}
   \sqrt{s} \gg M_D \hspace{0.5cm} \Rightarrow \hspace{0.5cm} 
   R_{\rm S} \gg \lambda_{\rm Pl}, \gg \lambda_{\rm B},  
\end{equation}
where
\begin{equation}
   \lambda_{\rm Pl} = G_D\,^{\frac{1}{\delta + 2}}
\end{equation}
is the Planck length in $D = (4 + \delta)$-dimensions, 
$\lambda_{\rm B} = \hbar / \sqrt{s}$ the de Broglie wavelenth, and 
\begin{equation}
   R_{\rm S}\, (\sqrt{s}) = \frac{1}{M_D} \left[\frac{\sqrt{s}}{M_D} 
   \left(\frac{2^{\delta} \pi^{\frac{\delta - 3}{2}} \Gamma
   \left(\frac{3 + \delta}{2} \right)}{2 + \delta} \right)
   \right]^{\frac{1}{1 + \delta}}, 
\label{Schwarzschild-radius} 
\end{equation}
is the Schwarzschild radius associated with the centre-of-mass energy
$\sqrt{s}$.
Higher-dimensional Schwarzschild black hole of radius 
$R_{\rm S}\, (M_{\rm bh})$ was originally obtained as a solution to the 
vacuum Einstein equations \cite{Myers:1986un}.
In extra dimension models containing branes, this solution is valid 
if one assumes that the brane tension does not perturb it strongly. 
A Schwarzschild black hole is $(4 + \delta)$--dimensional if its radius, 
eq.~(\ref{Schwarzschild-radius}), is small compared to the compactification 
radius of the extra dimenions, $R_{\rm S} \ll r_c$.
Since in the large extra dimension scenario the above conditions 
are always fulfilled, we shall restrict our further discussion to this 
scenario.

Phenomenology of trans-Planckian energy scattering in large extra dimension 
scenarios has been studied in Ref.~\cite{Giudice:2001ce,Emparan:2002kf},
which focus on the regime of large impact parameter $b \gg R_S$, where the 
elastic cross section is calculable using the eikonal approximation.
On the other hand, in the regime where black hole formation is conjectured,
\begin{equation}
   \sqrt{s} \gg M_D\, , \hspace{0.4cm} b < R_{\rm S}\, ,
\end{equation}
exact calculations are impossible due to the high non-linearity of the 
Einstein equations. 
Nevertheless, a geometrical parametrisation for the black hole production 
cross section at the parton-level $i j$,
\begin{equation}
  \sigma^{\rm bh}_{i j} ( s ) 
  \approx \pi R^2_{\rm S} \left(M_{\rm bh} = \sqrt{s} 
  \right)~\Theta \left(\sqrt{s} - 
  M^{\rm min}_{\rm bh} \right),
\label{bhp-cs}
\end{equation}
is believed to capture the essential features of this nonperturbative 
phenomenon
\cite{Banks:1999gd,Giddings:2002bu,Dimopoulos:2001hw,Eardley:2002re,Kohlprath:2002yh,Dimopoulos:2002qe} (see also Ref.~\cite{Jevicki:2002fq}). 
This semiclassical description is assumed to be valid above a minimum black 
hole mass $M^{\rm min}_{\rm bh} \gg M_D$, which is taken to be a free 
parameter besides $M_D$ and $\delta = D - 4$.  
Note however that there is still an ongoing debate on whether the 
production cross section for large black holes, i.e. black holes with 
masses $M_{\rm bh} \gg M_D$, is exponentially suppressed, as suggested by 
Voloshin \cite{Voloshin:2001vs}.

Let us end this subsection with some remarks concerning the working 
hypothesis (\ref{bhp-cs}). 
\begin{itemize}

\item 
In Ref.~\cite{Dimopoulos:2002qe} this problem is studied in string theory 
by applying the correspondence principle 
\cite{Horowitz:1997nw},
according to which a transition between black holes and highly excited 
long strings --``string balls'' occurs at the correspondence point
$M^{\rm min}_{\rm bh} \sim M_{\rm str} / g^2_{\rm str}$.
By assuming a string mass scale of $M_{\rm str} \sim$ TeV and working 
in the weak-coupling limits $g^2_{\rm str} \ll 1$, where there exists a 
broad stringy regime between the scales $M_{\rm str}$ and 
$M^{\rm min}_{\rm bh}$, $M^{\rm min}_{\rm bh} \gg M_D \gg M_{\rm str}$,
the cross section for string ball production is calculated and it matches 
the geometric cross section for black hole production at the 
correspondence point.

\item 
For simplicity the mass of the black hole is taken to be equal to the 
inital centre-of-mass energy $\sqrt{s}$, though part of it should escape 
to infinity in the form of gravitational radiation.
Penrose 
\cite{Penrose:1974}
has studied the collision of two black holes moving at nearly the speed of 
light using the approach of colliding plane gravitational shock waves
\cite{Aichelburg:1971dh}, 
and found an apparent horizon for a range of impact parameters.
By making use of the cosmic censorship conjecture he derived a lower bound 
on the mass of the final Schwarzschild black hole as 
$M_{\rm bh} > \sqrt{s /2}$ for head-on collision.
D'Eath and Payne \cite{D'Eath:1992hb}
obtained an improved estimate as $M_{\rm bh} \approx 0.84\, \sqrt{s}$ 
from their mass-loss formula
(for estimates obtained by other approaches, see e.g. 
Ref.~\cite{Cardoso:2002ay}).
Recently, Eardley and Giddings
\cite{Eardley:2002re}
redo this work for head-on collisions in $D \geq 4$ and non-zero impact 
parameters $b \neq 0$ in $D = 4$ dimensions using the same approach.
By finding a marginally trapped surface ${\cal S}$ in $D = 4$ and by 
applying the area theorem, they derived a lower bound on the mass of the 
final black hole as a function of the impact parameter $b$, ranging from
$0.71\, \sqrt{s}$ for $b = 0$ to $0.45\, \sqrt{s}$ for
$b = b_{\rm max} \approx 3.219\, G_{\rm N}\, ( \sqrt{s} / 2)$.
Lower limit on $M_{\rm bh}$ in head-on collisions $b = 0$ for various
dimensions $D$ were also obtained.
But corresponding estimate for the case $D > 4, b \neq 0$ is not available 
so far.

\item 
Generally speaking, one expects spinning black holes to form in 
particle collisions with nonzero impact parameters
(see Ref.~\cite{Park:2001xc,Anchordoqui:2001cg} for a first attempt to 
include this effect).
The higher-dimensional generalisation of the Kerr solution obtained in 
Ref.~\cite{Myers:1986un}
contains $[(D - 1) / 2]$ angular momentum parameters $J_i$,
and it is found that the properties of higher-dimensional rotating
black holes are essentially different from the four-dimensional ones.
In the case of the production of the $(4 + \delta)$-dimensional black holes
from the particle collisions on the 3-brane, only one angular momentum mode 
$J$ is excited.
But it is still not clear whether one can simply insert the Kerr radius 
determined from
\begin{equation}
   r_k \left[ 1 + r_k^{-2} \cdot 
   \frac{(\delta + 2)^2~J^2}{4 M^2_{\rm bh}} \right]^
   {\frac{1}{\delta + 1}} = R_{\rm S}
\end{equation}
into the geometric cross section, eq.~(\ref{bhp-cs}), instead of 
$R_{\rm S}$.

\item
In general one also expects black holes to inherit the electric charges 
from the initial state partons.
In Ref.~\cite{Casadio:2001wh} one finds a discussion on the charge of 
black holes and naked singularities produced in particle collisions 
in the warped scenario. 
But a generalisation of the Reissner-Nordstroem solution in the 
large extra dimension scenario is not available yet.

\item
It is proposed that in string theories, trans-Planckian energy scattering 
should generally lead to the creation of higher-dimensional, 
non-perturbative gravitational objects ($p$-branes), since spherically 
symmetric black holes are regarded as one particular class of them 
($0$-branes).
In Ref.~\cite{Ahn:2002mj} it is found that in some cases, $p$-brane formation 
could be competitive with black hole production, and has important 
phenomenological consequences \cite{Jain:2002kf}.

\end{itemize}

\subsection{Black hole decay}

In general, after its formation in high energy particle collisions, 
a black hole loses electric charge via the Schwinger process 
\cite{Schwinger:1951nm}
and angular momentum via superradiance \cite{Zel'dovich:1970} rapidly.
These processes are accompanied by the Hawking effect
\cite{Hawking:1974rv}, which reduces the black hole mass and hence 
its surface area.

As the black hole evaporates down to a mass of order of the Planck scale, 
one must again appeal to the theory of quantum gravity. 
It is also for this reason still not clear what is left over after the 
decay of a black hole.
Within string theory, a black hole evaporates only down to the 
correspondence point $M^{\rm min}_{\rm bh} \sim M_{\rm str} / g^2_{\rm str}$.
It then makes a transition to a ``string ball'', as mentioned in section 4.1.
Ref.~\cite{Dimopoulos:2002qe} 
gives a detailed description of the string ball decay processes.

But since a black hole spends most of its lifetime in the stage where 
its mass is close to the initial value, it is sufficient to adopt the 
following semiclassical approximation for the purpose of e.g. estimating 
black hole event rates and event signatures at colliders or at 
cosmic ray facilities, which will be illustrated in the next section.
For investigations of black hole evaporation in extra dimensions, 
see e.g. Ref.~\cite{Giddings:2002bu,Casadio:2001wh,Casadio:2000py,Kanti:2002nr,Frolov:2002as}.

Neglecting the backreaction of the emitted particles on the spacetime 
geometry (described by the greybody factor), a $(4 + \delta)$-dimensional 
Schwarzschild black hole of initial mass $M_{\rm bh} \gg M_D$
radiates thermally, 
as a black body with a surface area 
${\cal A}_{\delta + 2}$ 
at the Hawking temperature $T_{\rm H}$, which are
\begin{eqnarray}
   {\cal A}_{\delta + 2} &=& R\,^{\delta + 2}_{\rm S} \cdot 
   \Omega_{\delta + 2}\, , 
   \hspace{0.7cm} {\rm with} \hspace{0.3cm}
   \Omega_{\delta + 2} = \frac{2 \pi^{(\delta + 3) / 2}}
   {\Gamma (\frac{\delta + 3}{2})}, \nonumber \\
   T_{\rm H} &=& \frac{\delta + 1}{4 \pi R_{\rm S}}\, ,
\end{eqnarray}
in $(4 + \delta)$-dimensions. 
It is shown in Ref.~\cite{Emparan:2000rs} that the multi-dimensional 
black holes localised on the brane radiate at equal rates
\begin{equation}
   \frac{d E}{d t} \simeq \sigma_{\delta + 4}\, 
   {\cal A}_{\delta + 2}\, T\,^{\delta + 4}_{\rm H} \propto 
   \frac{1}{R^2_{\rm S}}\, ,
\end{equation}
into a bulk field and into a brane field (the Stefan-Boltzmann constant in 
$(\delta + 4)$-dimensions $\sigma_{\delta + 4}$ is found to be almost 
independent of the number of extra dimensions $\delta$). 
The fact that there are much more fields on the brane than in the bulk then
leads to the conclusion that small black holes localised on the brane  
radiate mainly into Standard Model particles on the brane rather than 
into the bulk. 
This raises the possibility of examining the black hole formation conjecture 
in future high-energy experiments via the observation of Hawking radiation,
provided TeV-scale gravity is realised in nature.

Total number of particles emitted is
characteristic of the entropy of the initial black hole
\begin{equation}
   S_{\rm bh} = \frac{{\cal A}_{\delta + 2}}{4\, G_D}\, . 
\end{equation}
During its lifetime,
\begin{equation}
   \tau_{D} \sim \sigma_{\delta + 4}\, \frac{1}{M_D} 
   \left(\frac{M_{\rm bh}}{M_D} \right)^{\frac{\delta + 3}
   {\delta + 1}} \sim 10^{-26}~{\rm s}\, \hspace{0.2cm} \left(
   \gg M^{-1}_{\rm bh} \right)\, ,
\end{equation} 
(for an average mini black hole of $M_{\rm bh} \sim {\cal O} (10)$ TeV), 
a mini black hole emits approximately 
\begin{equation}
   \left< n \right> \approx \frac{M_{\rm bh}}{2\, T_{\rm H}}   
\end{equation}
particles \cite{Dimopoulos:2001hw}, mostly hadrons and leptons.

\section{Search for black hole events}

Based on the expectation for production and evaporation of 
microscopic black holes in TeV-scale gravity scenarios outlined
in the last section, 
we want to report in this section on recent investigations of 
the corresponding phenomenology
at future colliders and cosmic ray facilities.

\subsection{Microscopic black hole production at colliders}

If TeV-scale gravity is realised in nature, the Large Hadron Collider 
(LHC), with its design values $\sqrt{s} = 14$ TeV for the proton-proton 
cm energy and ${\cal L} = 10^{34}~{\rm cm}^{-2}~{\rm s}^{-1}$
for the luminosity, will be producing black holes copiously at a rate
$d {\cal N}_{\rm bh} / d t = \sigma^{\rm bh}_{pp} \cdot {\cal L}$, with 
\begin{equation}
   \sigma^{\rm bh}_{pp} (s) = \sum_{i j} \int^1_0 d x_1 d x_2\,
   \frac{f_i (x_1,\mu) f_j (x_2, \mu) + f_i (x_2, \mu)
   f_j (x_1, \mu)}{1 + \delta_{i j}}\, \hat{\sigma}^{\rm bh}_{i j}
   (x_1 x_2 s)\, ,  
\end{equation}
the contribution of black hole production to the proton-proton cross
section.
The sum extends over all partons in the nucleon, with parton distributions 
$f_i (x, \mu)$ and factorisation scale $\mu$.
For $\hat{\sigma}^{\rm bh}_{i j}$, the black hole production cross section 
at the parton level, the geometric parametrisation (\ref{bhp-cs}) 
is usually adopted.    
 
Black hole detection at the LHC has been investigated in detail  
\cite{Giddings:2002bu,Dimopoulos:2001hw,LHC:all}.
First event simulations \cite{Dimopoulos:2001hw} show that the expected 
signature of black hole events at the LHC is the production of a number of 
${\cal O} (20)$ hard quanta, with energies approaching several hundreds 
of GeV.  
A substantial fraction of the beam energy is deposited in visible transverse 
energy, in an event with high sphericity.
Therefore, only a handful of such events is needed at the LHC to discriminate 
it against perturbative Standard Model background.

\begin{figure}[!ht]
\begin{center}
\includegraphics*[width=13cm]{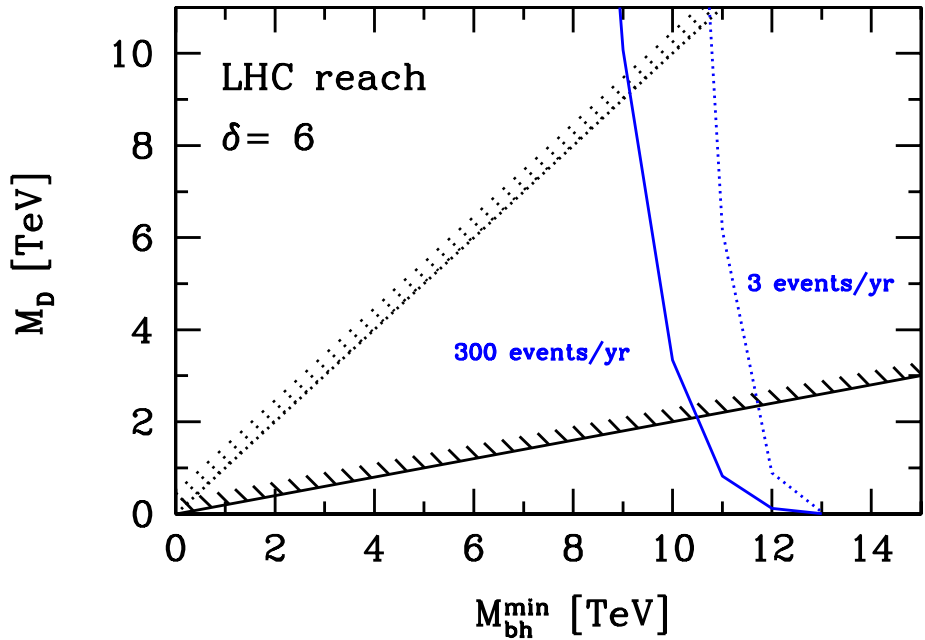}
\vspace{-1.8cm}
\caption[dum]{\small
Accessible region in the black hole production parameters at the LHC
for $\delta = 6$ extra dimensions.
The solid and the dotted lines are contours of constant numbers of 
produced black holes per year of masses larger than 
$M^{\rm min}_{\rm bh}$ by a fundamental Planck scale $M_D$.
The shaded dotted, $M_D = M^{\rm min}_{\rm bh}$, and shaded solid,
$M_D = (1 /5) M^{\rm min}_{\rm bh}$, lines give a rough indication of
a boundary of applicability of the semiclassical picture
\cite{Giddings:2002bu}.
}
\label{LHC}
\end{center}
\end{figure}

However, as can be seen in Fig.~\ref{LHC}, the LHC will become a black hole 
factory only if $M_D$ is below 2 TeV. 
In this case, not only the black hole formation conjecture and the existence 
of large extra dimensions can be verified, but also the higher-dimensional 
Hawking evaporation law can be tested from the correlation between 
the black hole mass and the black hole temperature deduced from the energy 
spectrum of its decay products \cite{Dimopoulos:2001hw}. 
Otherwise, the non-observation of black hole events at the LHC will not give 
more stringent limits on the large extra dimension model than those to be 
derived from the searches for the manifestations of Kaluza-Klein gravitons 
listed in Table~\ref{LED}.

\subsection{Microscopic black holes from ultrahigh energy cosmic
neutrinos} 

Particle astrophysics experiments have demonstrated their complementarity 
to collider searches for New Physics in many aspects. 
In the case of black hole production in TeV-scale gravity, cosmic ray 
facilities such as Pierre Auger Observatory and neutrino telescopes such 
as AMANDA/IceCube and RICE also have an opportunity to see the first sign 
or place sensible constraints before LHC starts operating in 2007.

\begin{figure}[!ht]
\begin{center}
\includegraphics*[width=13cm]{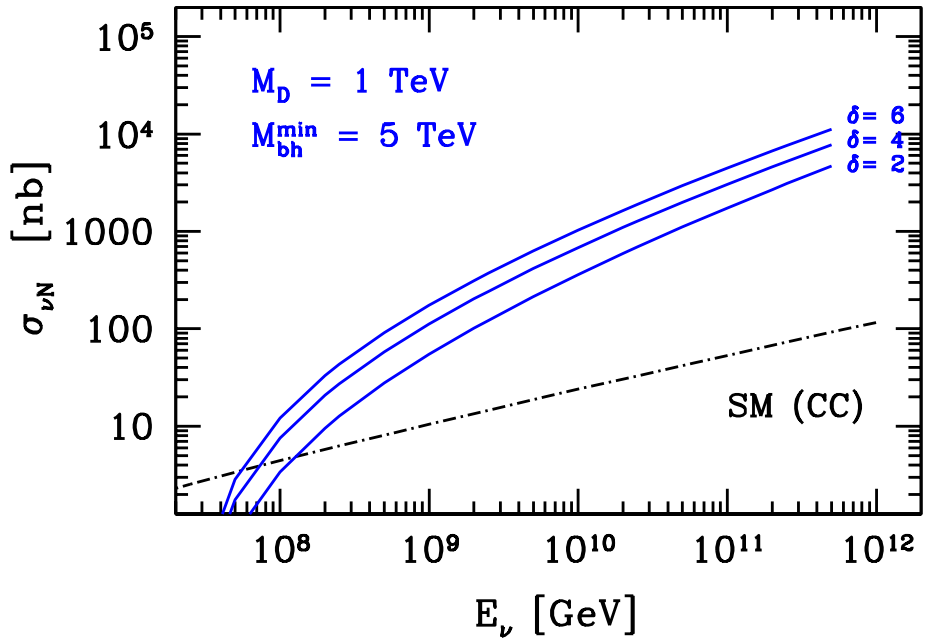}
\vspace{-1.8cm}
\caption[dum]{\small
Cross section $\sigma^{\rm bh}_{\nu N}$, eq.~(\ref{BH-sig-nuN}),
for black hole production in neutrino-nucleon scattering, as a function 
of the incident neutrino energy $E_{\nu}$.
Solid lines from bottom to top correspond to $\delta = 2, 4, 6$
large extra dimensions and $M_D = 1~{\rm TeV}$, 
$M^{\rm min}_{\rm bh} = 5~{\rm TeV}$. 
The Standard Model (SM) charged-current (CC) neutrino-nucleon cross section 
is also shown (dashed-dotted line).   
}
\label{BH-SM}
\end{center}
\end{figure}

The most distinct signature for cosmic ray facilities to look for are 
quasi-horizontal air showers, initiated by black holes produced in the 
scattering of ultrahigh energy cosmic neutrinos on nucleons in the 
atmosphere \cite{Feng:2002ib,Anchordoqui:2002ei,Ringwald:2002vk}.
Neutrinos interact only weakly in the Standard Model so an enhancement
in the total cross section due to the contribution from black hole 
production 
\begin{equation}
   \sigma^{\rm bh}_{\nu N} (s) = \sum_i \int^1_0 dx~f_i (x, \mu)
   \hat{\sigma}^{\rm bh}_{\nu N} (x s)
\label{BH-sig-nuN}
\end{equation}
would be quite noticeable (see Fig.~\ref{BH-SM}).

A cosmic ray facility of acceptance ${\cal A}$ would expect such events 
at a rate (neglecting the neutrino flux attenuation in the upper atmosphere)
\begin{equation}
   \frac{d {\cal N}^{\rm bh}_{\rm sh}}{d t} 
   (> E_{\rm th}) =    
   N_A~\rho_{\rm air} \int^{\infty}_{E_{\rm th}}
   d E_{\nu}~F_{\nu} (E_{\nu}) \sigma^{\rm bh}_{\nu N} (E_{\nu})
   {\cal A} (E_{\nu})\, ,
\label{bh-auger}
\end{equation}
where $N_A$ is the Avogadro's constant and 
$\rho_{\rm air} \simeq 10^{-3}~{\rm g}~{\rm cm}^{-3}$ the air density. 
Equation (\ref{bh-auger}) assumes that 100\% of the incident neutrino
energy goes into visible, hadronic or electromagnetic shower energy. 
But most uncertainties come from the unknown ultrahigh energy neutrino
fluxes $F_{\nu} (E_{\nu})$.

\begin{figure}[!ht]
\begin{center}
\includegraphics*[width=13cm]{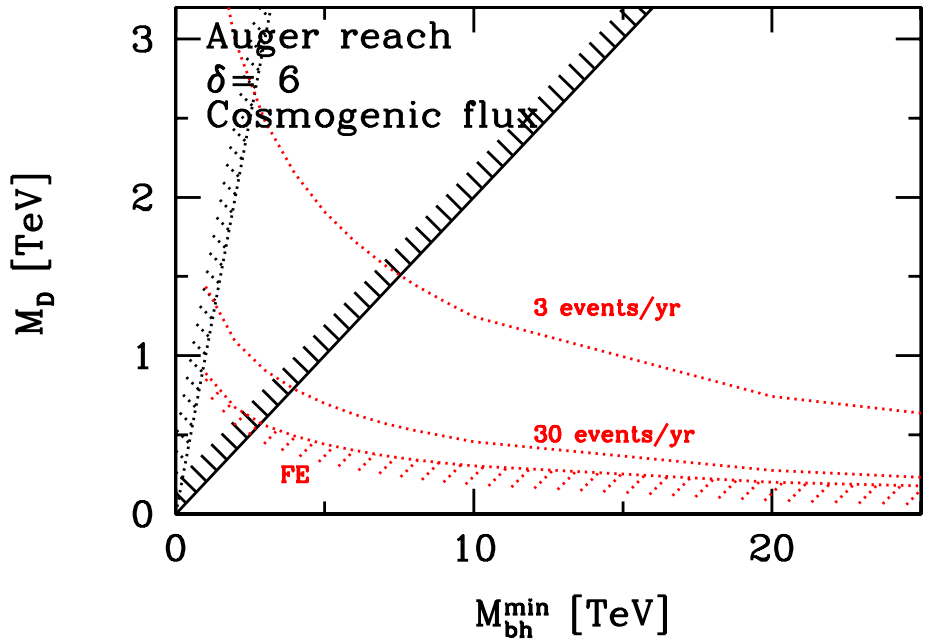}
\vspace{-1.8cm}
\caption[dum]{\small
Projected Auger reach in the black hole production parameters for
$\delta = 6$ large extra dimensions, by exploiting the cosmogenic neutrino 
flux from Ref.~\cite{Protheroe:1996ft} with cutoff energy 
$3 \times 10^{21}$ eV for the ultrahigh energy cosmic rays.
The shaded dotted, $M_D = M^{\rm min}_{\rm bh}$, and shaded solid,
$M_D = (1 / 5) M^{\rm min}_{\rm bh}$, lines give a rough indication of the
boundary of applicability of the semiclassical picture  
\cite{Giddings:2002bu}.
The dotted lines respresent the contour of 3 and 30, respectively,
detected horizontal air showers per year initiated by neutrino-nucleon
scattering into a black hole of mass larger than $M^{\rm min}_{\rm bh}$,
for a fundamental Planck scale $M_D$.
Also shown is the constraint arising from the non-observation of horizontal 
air showers by the Fly's Eye collaboration (shaded dotted line labeled 
``FE'' \protect\cite{Baltrusaitis:mt}).
The constraint imposed by AGASA, first obtained in 
Ref.~\protect\cite{Anchordoqui:2001cg}, is slightly above the 
30 events/yr contour line for Auger.  
}
\label{bh_auger_pj}
\end{center}
\end{figure}

Cosmic neutrinos of energies higher than $10^7$ GeV from various sources 
are predicted (for recent reviews, see Ref.~\cite{Nu:review}).
Among all, the cosmogenic ones expected from cosmic ray 
interactions with CMBR
(e.g. $p \gamma \rightarrow \Delta \rightarrow n \pi^+ 
\rightarrow \nu_{\mu} \bar{\nu}_{\mu} \nu_e...$)
are more or less guaranteed to exist.
By exploiting the cosmogenic neutrino fluxes
(recent estimates for these fluxes can be found in Ref.
\cite{Protheroe:1996ft,Cosmogenic})
one thus obtains a conservative lower limit on the projected Auger
sensitivity to black hole production
(Fig.~\ref{bh_auger_pj}).

In Ref.~\cite{Anchordoqui:2002ei} it was shown that these showers may have 
an ``anomalous'' electromagnetic component: about one order of magnitude 
larger than Standard Model $\nu_{\mu}$-initiated showers and at least an 
order of magnitude smaller than Standard Model $\nu_e$-initiated showers.
It was argued that this represents a very clean signal and, correspondingly, 
that for the ground array of the Auger Observatory a total number of about 
${\cal O} (20)$ black hole events could give significant statistics for a 
discrimination against the signal from the Standard Model background, 
which is estimated to be $\sim 0.15$ events/yr assuming the same cosmogenic 
neutrino flux \cite{Protheroe:1996ft}.

Somewhat later it is argued in Ref.~\cite{Anchordoqui:2001cg} that, 
based on the comparison with the earth-skimming neutrino events, 
an excess of a handful of quasi-horizontal black hole events are 
sufficient for a discrimination against the Standard Model background.
An inspection of Fig.~\ref{bh_auger_pj} thus leads to the conclusion
that, already for an ultrahigh energy neutrino flux at the cosmogenic level 
estimated in Ref.~\cite{Protheroe:1996ft}, the Pierre Auger Observatory, 
expected to become fully operational in 2003, has the opportunity to 
see first signs of black hole production. 

On the other hand, the non-observation of horizontal air showers reported 
by the Fly's Eye and the AGASA collaboration provides a stringent bound on 
$M_D$, which is competitive with existing bounds on $M_D$ from colliders 
as well as from astrophysical and cosmological considerations listed in 
Table.~\ref{LED}, particularly for larger numbers of extra dimensions 
($\delta \geq 5$) and smaller threshold 
($M^{\rm min}_{\rm bh} \lesssim 10$ TeV) for the semiclassical description, 
eq.~(\ref{bhp-cs}).

The sensitivity of neutrino telescopes such as AMANDA/IceCube, ANTARES, 
Baikal, NESTOR and RICE to black hole production in the scattering of 
ultrahigh energy cosmic neutrinos on nucleons in the ice or water has been 
investigated in Ref.~\cite{Kowalski:2002gb} and compared to the one 
expected at the cosmic ray facilities and at the LHC.
The underwater/-ice neutrino telescopes are sensitive not only to the 
contained black hole events but also to the through-going muons produced 
in the black hole decay outside the detector.  

\begin{figure} 
\begin{center}
\includegraphics*[bbllx=20pt,bblly=221pt,bburx=585pt,bbury=608pt,%
width=13cm]{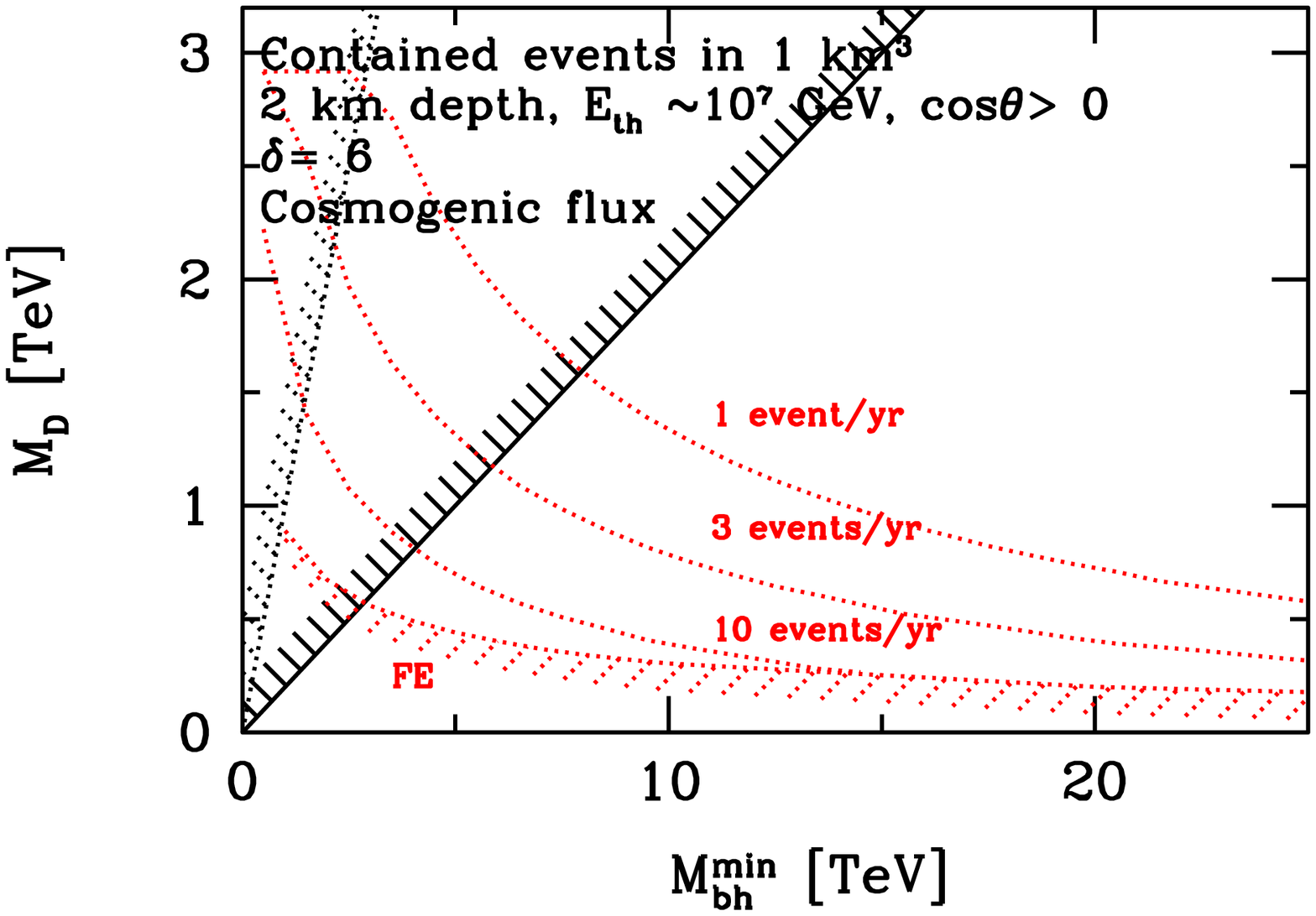}
\hspace{1cm}
\includegraphics*[bbllx=20pt,bblly=221pt,bburx=585pt,bbury=608pt,%
width=13cm]{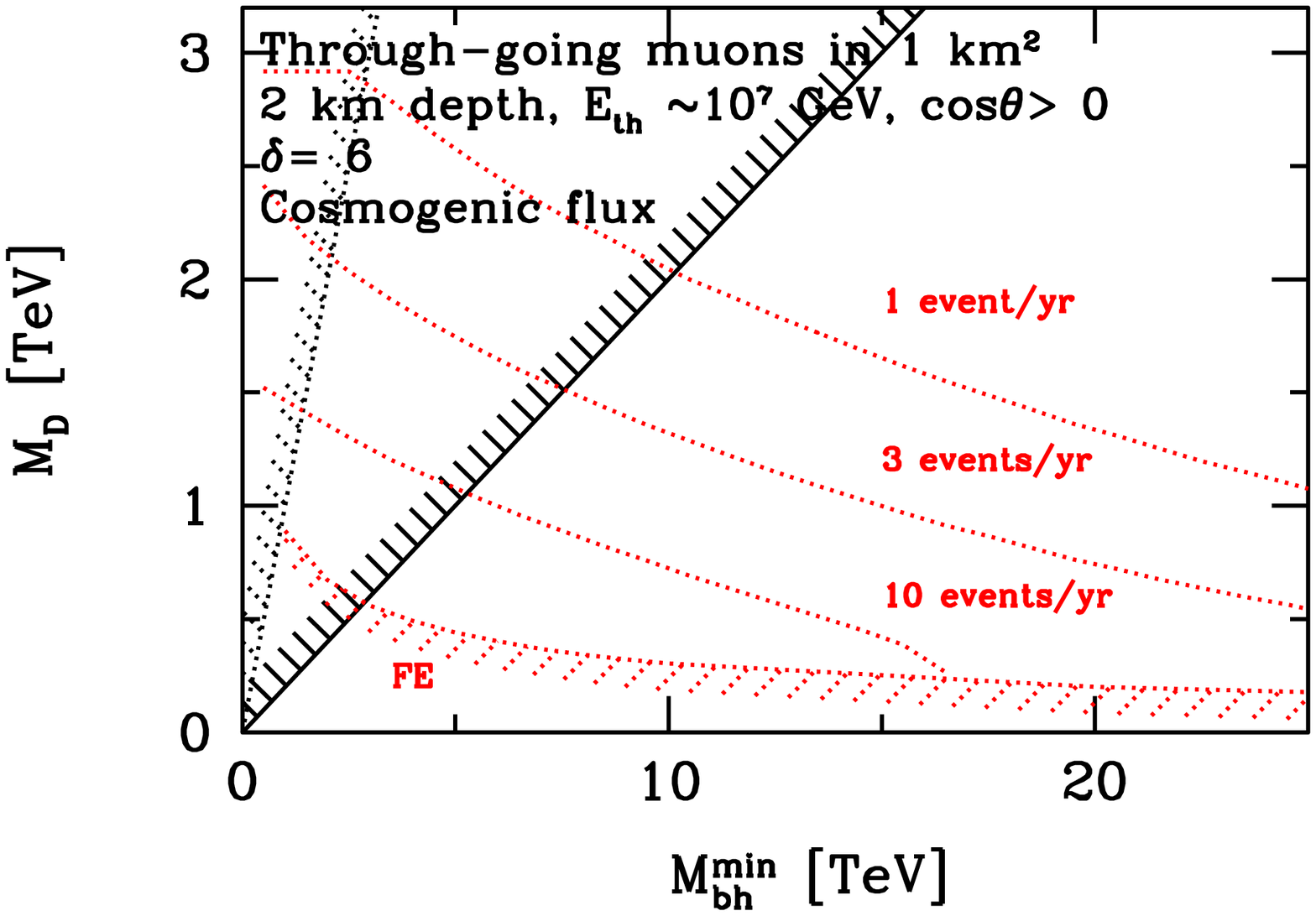}
\caption[dum]{\small
Reach of the neutrino telescopes in the black hole production parameters 
for $\delta = 6$ large extra dimensions, with the shaded dotted, 
$M_D = M^{\rm min}_{\rm bh}$, shaded solid, $M_D = (1 / 5) 
M^{\rm min}_{\rm bh}$, lines and the shaded dotted line labeled ``FE'' 
same as in Fig.~\ref{bh_auger_pj}.
Top: for contained events in an under-ice detector at a depth of 2 km 
and with an 1 km$^3$ fiducial volume.
Bottom: for through-going muons in an under-ice detector at a depth of 2 km 
and with an 1 km$^2$ effective area.
Both by exploiting the cosmogenic neutrino flux from 
Ref.~\cite{Protheroe:1996ft} with cutoff energy $3 \times 10^{21}$ eV 
for the ultrahigh energy cosmic rays.
}
\label{amanda-casc-pj}
\end{center}
\end{figure}

The projected reach in the black hole production parameter space for 
contained events in an under-ice detector corresponding to the IceCube 
porposal (2 km depth, fiducial volume 1 km$^3$), as well as for through-going 
muons in an under-ice detector at a depth of 2 km and with an effective area 
of 1 km$^2$, are shown in Fig.~\ref{amanda-casc-pj}.
A cosmogenic neutrino flux from Ref.~\cite{Protheroe:1996ft} is again assumed. 
Taking into account the small effective volume, 
$V \approx 0.001 - 0.01 {\rm km}^3$, of the currently operating AMANDA 
and Baikal neutrino telescopes and the time schedule of IceCube
(expects to reach the final effective volume $V \approx 1 {\rm km}^3$ 
only after the start of the LHC), the best perspective for black hole 
detection on the basis of the contained events must be assigned to RICE, 
a currently operating radio-Cherenkov neutrino detector with an effective
volume of $\approx 1 {\rm km}^3$ for $10^8$ GeV electromagnetic cascades.
Using already available data, RICE could set sensible constraints on 
black hole production parameters.

The ability to detect muons from distant neutrino reactions increases an
underwater/-ice detector's effective neutrino target volume dramatically.
With an effective area of about 0.3 km$^2$ for down-going muons
above 10$^7$ GeV and 5 years data available, AMANDA achieves an exposure 
of about $1 {\rm km}^2~{\rm yr}$. 
In the case that the neutrino flux is just at the level of the
cosmogenic one, only a few $(\lwig 1)$ through-going events from 
Standard Model background are expected per year.  
Therefore, AMANDA should be able to impose strong constraints on the 
black hole production parameters if the currently available data show no 
through-going muons above 10$^7$ GeV. 
Furthermore, if an ultrahigh energy cosmic neutrino flux significantly 
higher than the cosmogenic one is realised in nautre,
one even has discovery potential for black holes at IceCube beyond the 
reach of LHC, though discrimination between Standard Model background
and black hole events becomes crucial. 
For other quite similar investigations on the sensitivity of neutrino 
telescopes to black hole production, see Ref.~\cite{Alvarez-Muniz:2002ga}.

\section{Conclusion}

Models with extra spatial dimensions and TeV-scale gravity offer the first 
opportunity to test the conjecture of black hole formation in 
trans-Planckian energy collisions and the prediction of black hole 
evaporation by near-future high energy experiments.   
The LHC will be producing microscopic black holes copiously if the 
fundamental Planck scale $M_D$ is below 2 TeV, 
while the reach of the cosmic ray facilities and neutrino telescopes depends 
on the unknown ultrahigh energy cosmic neutrino flux.

It is found that, already for a cosmic neutrino flux at the level expected 
from the cosmic ray primary interactions with the cosmic microwave
background radiation, the Pierre Auger Observatory has an opportunity 
to see first signs of black hole production before the start of the LHC.
The non-observation of horizontal air showers by the Fly's Eye and AGASA
experiments imposes a stringent bound on $M_D$, which is competitive with
existing bounds from colliders as well as from astrophysical and cosmogolical
considerations. 
Further sensible constraints on the black hole production parameters and/or
bounds on low-scale gravity are expected from neutrino detectors AMANDA 
and RICE, using already available data for through-going muons and 
contained events, respectively.
Moreover, in the optimistic case that an ultrahigh energy cosmic neutrino 
flux is significantly higher than the cosmogenic one, one even has 
discovery potential for black holes at the under-ice neutrino detector 
IceCube beyond the reach of LHC.

We have illustrated the search strategies at colliders, cosmic ray
facilities and neutrino telescopes, and compared their reaches
with the simplest, nevertheless representative example:
production of Schwarzschild black holes in large extra dimension  
scenarios.
There still remains much work to be done. 

On the theoretical side, before the full quantum theory of gravity is 
developed, one can still improve/extend the estimate of the 
black hole production cross section in $D \geq 4$ dimensions 
semiclassically. 
Other tasks not necessarily requiring the knowledge of quantum gravity 
include the construction of higher-dimensional black hole solutions 
with charges and angular momenta in different models,
and evaluating the grey-body factor for the decay of the 
higher-dimensional black holes.

While the above theoretical refinements may not affect the 
expected black hole signatures very much,   
more simulations for various detectors need to be done, 
in order to develop successful tools to
discriminate black hole events 
against Standard Model backgrounds.


\section*{Acknowledgements}
We thank R.~Casadio, R.~Emparan, J.~Feng and A.~Ringwald
for helpful discussions.


\begin{thebibliography}{99}



\bibitem{'tHooft:1987rb}
G.~'t Hooft,
Phys.\ Lett.\ B {\bf 198} (1987) 61;
%
Nucl.\ Phys.\ B {\bf 304} (1988) 867.
%

\bibitem{Amati:1987wq}
D.~Amati, M.~Ciafaloni and G.~Veneziano,
Phys.\ Lett.\ B {\bf 197} (1987) 81;
%
Int.\ J.\ Mod.\ Phys.\ A {\bf 3} (1988) 1615;
%
Phys.\ Lett.\ B {\bf 216} (1989) 41;
%
Nucl.\ Phys.\ B {\bf 347} (1990) 550;
%
Phys.\ Lett.\ B {289} (1992) 87;
%
Nucl.\ Phys.\ B {403} (1993) 707.
%

\bibitem{Kabat:1992tb}
D.~Kabat and M.~Ortiz,
Nucl.\ Phys.\ B {\bf 388} (1992) 570
[arXiv:hep-th/9203082].
%

\bibitem{Verlinde:1992iu}
H.~Verlinde and E.~Verlinde,
Nucl.\ Phys.\ B {\bf 371} (1992) 246
[arXiv:hep-th/9110017].
%

\bibitem{Fabbrichesi:1994kd}
M.~Fabbrichesi, R.~Pettorino, G.~Veneziano and G.~A.~Vilkovisky,
Nucl.\ Phys.\ B {\bf 419} (1994) 147.
%

\bibitem{Aref'eva:1995qs}
I.~Y.~Aref'eva, K.~S.~Viswanathan and I.~V.~Volovich,
Nucl.\ Phys.\ B {\bf 452} (1995) 346
[Erratum-ibid.\ B {\bf 462} (1995) 613];
%
Int.\ J.\ Mod.\ Phys.\ D {\bf 5} (1996) 707.
%

\bibitem{Arkani-Hamed:1998rs}
N.~Arkani-Hamed, S.~Dimopoulos and G.~R.~Dvali,
Phys.\ Lett.\ B {\bf 429} (1998) 263
[arXiv:hep-ph/9803315];
%
I.~Antoniadis, N.~Arkani-Hamed, S.~Dimopoulos and G.~R.~Dvali,
Phys.\ Lett.\ B {\bf 436} (1998) 257
[arXiv:hep-ph/9804398].
%

\bibitem{Randall:1999ee}
L.~Randall and R.~Sundrum,
Phys.\ Rev.\ Lett.\  {\bf 83} (1999) 3370
[arXiv:hep-ph/9905221].
%
\bibitem{Argyres:1998qn}
P.~C.~Argyres, S.~Dimopoulos and J.~March-Russell,
Phys.\ Lett.\ B {\bf 441} (1998) 96
[arXiv:hep-th/9808138].
%

\bibitem{Banks:1999gd}
T.~Banks and W.~Fischler,
arXiv:hep-th/9906038.
%

\bibitem{Aref'eva:2000bm}
I.~Y.~Aref'eva,
Part.\ Nucl.\  {\bf 31} (2000) 169
[arXiv:hep-th/9910269].
%

\bibitem{Emparan:2000rs}
R.~Emparan, G.~T.~Horowitz and R.~C.~Myers,
Phys.\ Rev.\ Lett.\  {\bf 85} (2000) 499
[arXiv:hep-th/0003118].
%

\bibitem{Giddings:2001ay}
S.~B.~Giddings and E.~Katz,
J.\ Math.\ Phys.\  {\bf 42} (2001) 3082
[arXiv:hep-th/0009176].
%

\bibitem{Emparan:2001ce}
R.~Emparan,
Phys.\ Rev.\ D {\bf 64} (2001) 024025
[arXiv:hep-th/0104009].
%

\bibitem{Hawking:1974rv}
S.~W.~Hawking,
Nature {\bf 248} (1974) 30;
%
Commun.\ Math.\ Phys.\  {\bf 43} (1975) 199.
%

\bibitem{Giddings:2002bu}
S.~B.~Giddings and S.~Thomas,
Phys.\ Rev.\ D {\bf 65} (2002) 056010
[arXiv:hep-ph/0106219].
%

\bibitem{Dimopoulos:2001hw}
S.~Dimopoulos and G.~Landsberg,
Phys.\ Rev.\ Lett.\  {\bf 87} (2001) 161602
[arXiv:hep-ph/0106295].
%

\bibitem{Feng:2002ib}
J.~L.~Feng and A.~D.~Shapere,
Phys.\ Rev.\ Lett.\  {\bf 88} (2002) 021303
[arXiv:hep-ph/0109106];
%

\bibitem{Anchordoqui:2001cg}
L.~A.~Anchordoqui, J.~L.~Feng, H.~Goldberg and A.~D.~Shapere,
Phys.\ Rev.\ D {\bf 65} (2002) 124027
[arXiv:hep-ph/0112247];
%
arXiv:hep-ph/0207139.
%


\bibitem{Anchordoqui:2002ei}
L.~Anchordoqui and H.~Goldberg,
Phys.\ Rev.\ D {\bf 65} (2002) 047502
[arXiv:hep-ph/0109242].
%

\bibitem{Uehara:2001yk}
Y.~Uehara,
arXiv:hep-ph/0110382.
%

\bibitem{Ringwald:2002vk}
A.~Ringwald and H.~Tu,
Phys.\ Lett.\ B {\bf 525} (2002) 135
[arXiv:hep-ph/0111042].
%

\bibitem{Kowalski:2002gb}
M.~Kowalski, A.~Ringwald and H.~Tu,
Phys.\ Lett.\ B {\bf 529} (2002) 1
[arXiv:hep-ph/0201139].
%


\bibitem{Alvarez-Muniz:2002ga}
J.~Alvarez-Muniz, J.~L.~Feng, F.~Halzen, T.~Han and D.~Hooper,
Phys.\ Rev.\ D {\bf 65} (2002) 124015
[arXiv:hep-ph/0202081];
%
S.~I.~Dutta, M.~H.~Reno and I.~Sarcevic,
Phys.\ Rev.\ D {\bf 66} (2002) 033002
[arXiv:hep-ph/0204218].
%

\bibitem{Aichelburg:1971dh}
P.~C.~Aichelburg and R.~U.~Sexl,
Gen.\ Rel.\ Grav.\  {\bf 2} (1971) 303.
%

\bibitem{eikonal:1969}
H.~D.~Abarbanel and C.~Itzykson,
Phys.\ Rev.\ Lett.\  {\bf 23} (1969) 53;
%
M.~Levy and J.~Sucher,
Phys.\ Rev.\  {\bf 186} (1969) 1656.
%

\bibitem{Banados:1992wn}
M.~Banados, C.~Teitelboim and J.~Zanelli,
Phys.\ Rev.\ Lett.\  {\bf 69} (1992) 1849
[arXiv:hep-th/9204099];
%
M.~Banados, M.~Henneaux, C.~Teitelboim and J.~Zanelli,
Phys.\ Rev.\ D {\bf 48} (1993) 1506
[arXiv:gr-qc/9302012].
%

\bibitem{Matschull:1999rv}
H.~J.~Matschull,
Class.\ Quant.\ Grav.\  {\bf 16} (1999) 1069
[arXiv:gr-qc/9809087].
%

\bibitem{Penrose:1974}
R.~Penrose, presented at the Cambridge University Seminar,
Cambridge, England, 1974 (unpublished).
%


\bibitem{D'Eath:1992hb}
P.~D.~D'Eath and P.~N.~Payne,
Phys.\ Rev.\ D {\bf 46} (1992) 658;
%
Phys.\ Rev.\ D {\bf 46} (1992) 675;
%
Phys.\ Rev.\ D {\bf 46} (1992) 694.
%

\bibitem{D'Eath:1996nf}
P.~D.~D'Eath,
``Black holes: Gravitational interactions'',
{\it  Oxford, UK: Clarendon (1996)} 
%

\bibitem{Eardley:2002re}
D.~M.~Eardley and S.~B.~Giddings,
Phys.\ Rev.\ D {\bf 66} (2002) 044011
[arXiv:gr-qc/0201034].
%

\bibitem{Stark:it}
R.~F.~Stark and T.~Piran,
in {\it  *Kyoto 1986, Proceedings, Gravitational Collapse and 
Relativity* 249-275. }
%



\bibitem{Solodukhin:2002ui}
S.~N.~Solodukhin,
arXiv:hep-ph/0201248;
%
S.~D.~Hsu,
arXiv:hep-ph/0203154;
%
S.~Bilke, E.~Lipartia and M.~Maul,
arXiv:hep-ph/0204040.
%

\bibitem{Kohlprath:2002yh}
E.~Kohlprath and G.~Veneziano,
JHEP {\bf 0206} (2002) 057
[arXiv:gr-qc/0203093].
%



\bibitem{Rubakov:2001kp}
V.~A.~Rubakov,
Phys.\ Usp.\  {\bf 44} (2001) 871
[Usp.\ Fiz.\ Nauk {\bf 171} (2001) 913]
[arXiv:hep-ph/0104152].
%

\bibitem{Giudice:2001ce}
G.~F.~Giudice, R.~Rattazzi and J.~D.~Wells,
Nucl.\ Phys.\ B {\bf 630} (2002) 293
[arXiv:hep-ph/0112161].
%

\bibitem{Adelberger:2002ic}
E.~G.~Adelberger  [EOT-WASH Group Collaboration],
arXiv:hep-ex/0202008.
%

\bibitem{Abe:2001nq}
T.~Abe {\it et al.}  [American Linear Collider Working Group Collaboration],
hep-ex/0106057.


\bibitem{Hannestad:2002xi}
S.~Hannestad and G.~G.~Raffelt,
Phys.\ Rev.\ Lett.\  {\bf 88} (2002) 071301
[arXiv:hep-ph/0110067];
%
S.~Hannestad,
private communication. 
%


\bibitem{LED:limits}
M.~E.~Peskin,
hep-ph/0002041;
%
C.~Pagliarone,
hep-ex/0111063.
%

\bibitem{Myers:1986un}
R.~C.~Myers and M.~J.~Perry,
Annals Phys.\  {\bf 172} (1986) 304.
%

\bibitem{Emparan:2002kf}
R.~Emparan, M.~Masip and R.~Rattazzi,
Phys.\ Rev.\ D {\bf 65} (2002) 064023
[arXiv:hep-ph/0109287].
%

\bibitem{Dimopoulos:2002qe}
S.~Dimopoulos and R.~Emparan,
Phys.\ Lett.\ B {\bf 526} (2002) 393
[arXiv:hep-ph/0108060].
%

\bibitem{Jevicki:2002fq}
A.~Jevicki and J.~Thaler,
Phys.\ Rev.\ D {\bf 66} (2002) 024041
[arXiv:hep-th/0203172].
%


\bibitem{Voloshin:2001vs}
M.~B.~Voloshin,
Phys.\ Lett.\ B {\bf 518} (2001) 137
[arXiv:hep-ph/0107119];
%
Phys.\ Lett.\ B {\bf 524} (2002) 376
[arXiv:hep-ph/0111099].
%

\bibitem{Horowitz:1997nw}
G.~T.~Horowitz and J.~Polchinski,
Phys.\ Rev.\ D {\bf 55} (1997) 6189
[arXiv:hep-th/9612146].
%

\bibitem{Cardoso:2002ay}
V.~Cardoso and J.~P.~Lemos,
Phys.\ Lett.\ B {\bf 538} (2002) 1
[arXiv:gr-qc/0202019].
%

\bibitem{Park:2001xc}
S.~C.~Park and H.~S.~Song,
hep-ph/0111069.
%


\bibitem{Casadio:2001wh}
R.~Casadio and B.~Harms,
arXiv:hep-th/0110255.
%

\bibitem{Ahn:2002mj}
E.~J.~Ahn, M.~Cavaglia and A.~V.~Olinto,
arXiv:hep-th/0201042.
%

\bibitem{Jain:2002kf}
P.~Jain, S.~Kar, S.~Panda and J.~P.~Ralston,
arXiv:hep-ph/0201232;
%
L.~A.~Anchordoqui, J.~L.~Feng and H.~Goldberg,
Phys.\ Lett.\ B {\bf 535} (2002) 302
[arXiv:hep-ph/0202124];
%
K.~Cheung,
Phys.\ Rev.\ D {\bf 66} (2002) 036007
[arXiv:hep-ph/0205033].
%



\bibitem{Schwinger:1951nm}
J.~S.~Schwinger,
Phys.\ Rev.\  {\bf 82} (1951) 664.
%


\bibitem{Zel'dovich:1970}
Ya.~B.~Zel'dovich,
Pis'ma\ v\ Zh.\ Eksp.\ Theo.\ Fiz.\ {12} (1970) 443;
%
Sov.\ Phys.\ JHEP\ Lett.\ {14} (1971) 180;
%
Sov.\ Phys.\ JHEP\ Lett.\ {35} (1972) 1085;
%
C.~W.~Misner,
Phys.\ Rev.\ Lett.\  {\bf 28} (1972) 994.
%



\bibitem{Casadio:2000py}
R.~Casadio and B.~Harms,
Phys.\ Lett.\ B {\bf 487} (2000) 209
[arXiv:hep-th/0004004];
%
Phys.\ Rev.\ D {\bf 64} (2001) 024016
[arXiv:hep-th/0101154].
%

\bibitem{Kanti:2002nr}
P.~Kanti and J.~March-Russell,
Phys.\ Rev.\ D {\bf 66} (2002) 024023
[arXiv:hep-ph/0203223];
%
A.~V.~Kotwal and S.~Hofmann,
arXiv:hep-ph/0204117.
%

\bibitem{Frolov:2002as}
V.~Frolov and D.~Stojkovic,
Phys.\ Rev.\ D {\bf 66} (2002) 084002
[arXiv:hep-th/0206046];
%
Phys.\ Rev.\ Lett.\  {\bf 89} (2002) 151302
[arXiv:hep-th/0208102];
%
arXiv:gr-qc/0211055.
%




\bibitem{LHC:all}
S.~Hossenfelder, S.~Hofmann, M.~Bleicher and H.~Stocker,
hep-ph/0109085;
%
K.~m.~Cheung,
Phys.\ Rev.\ Lett.\  {\bf 88} (2002) 221602
[arXiv:hep-ph/0110163];
%
T.~G.~Rizzo,
hep-ph/0111230;
%
G.~Landsberg,
Phys.\ Rev.\ Lett.\  {\bf 88} (2002) 181801
[arXiv:hep-ph/0112061];
%
M.~Bleicher, S.~Hofmann, S.~Hossenfelder and H.~St\"ocker,
hep-ph/0112186.
%






\bibitem{Nu:review}
R.~J.~Protheroe,
Nucl.\ Phys.\ Proc.\ Suppl.\  {77} (1999) 465;
%
R.~Gandhi,
Nucl.\ Phys.\ Proc.\ Suppl.\  {91} (2000) 453;
%
J.~G.~Learned and K.~Mannheim,
Ann.\ Rev.\ Nucl.\ Part.\ Sci.\  {50} (2000) 679.
%

\bibitem{Protheroe:1996ft}
R.~J.~Protheroe and P.~A.~Johnson,
Astropart.\ Phys.\  {4} (1996) 253
[Erratum-ibid.\  {5} (1996) 215].
%


\bibitem{Cosmogenic}
S.~Yoshida and M.~Teshima,
Prog.\ Theor.\ Phys.\  {89} (1993) 833;
%
S.~Yoshida, H.~Dai, C.~C.~Jui and P.~Sommers,
Astrophys.\ J.\  {479} (1997) 547;
%
R.~Engel and T.~Stanev,
Phys.\ Rev.\ D {64} (2001) 093010.
%

\bibitem{Baltrusaitis:mt}
R.~M.~Baltrusaitis {\it et al.},
Phys.\ Rev.\ D {\bf 31} (1985) 2192.
%











\end{thebibliography}
\end{document}